# Decameter Type III Bursts with Changing Frequency Drift-Rate Signs

V.N. Melnik (1), A.I. Brazhenko (2), A.A. Konovalenko (1), C. Briand (3), V.V. Dorovskyy (1), P. Zarka (3), A.V. Frantsuzenko (2), H.O. Rucker (4), B.P. Rutkevych (1), M. Panchenko (4), L. Denis (3), T. Zaqarashvili (4), B. Shergelashvili (5)

(1) - Institute of Radio Astronomy, National Academy of Sciences of Ukraine, Chervonopraporna st. 4, 61002 Kharkiv, Ukraine

(2) - Institute of Geophysics, Myasoedova str. 27/29, 36014 Poltava, Ukraine

(3) - LESIA, UMR CNRS 8109, Observatoire de Paris, 92195 Meudon, France

(4) - Space Research Institute, Austrian Academy of Sciences, Schmiedlstrasse 6, 8042 Graz, Austria

(5) - Catholic University of Leuven, Celestijnenlaan 200B box 2400, 3001 Leuven, Belgium

**Abstract** We discuss properties of type III bursts that change the sign of their drift rate from negative to positive and vice versa. Moreover, these bursts may change the sign of their drift rates more than once. These particular type III bursts were observed simultaneously by the radio telescopes UTR-2 (Ukrainian T-shaped Radio telescope, Kharkov, Ukraine), URAN-2 (Ukrainian Radio telescope of the Academy of Sciences, Poltava, Ukraine), and by the NDA (Nançay Decametric Array, Nancay, France) in the frequency range 8 – 41 MHz. The negative drift rates of these bursts are similar to those of previously reported decameter type III bursts and vary from $-0.7$ MHz s$^{-1}$ to $-1.7$ MHz s$^{-1}$, but their positive drift rates vary in a wider range from 0.44 MHz s$^{-1}$ to 6 MHz s$^{-1}$. Unlike inverted U-bursts, the tracks of these type III bursts have C- or inverted C-shapes.

Our basic explanation of the positive drift rate of these type III bursts differs from the common assumption that positive drift rates of type III bursts are connected with electron beam propagation toward the Sun. We propose that, even if electron beams move outward from the Sun, they can generate type III bursts with positive drift rates if in some regions of the solar corona the group velocities of type III radio emissions are lower than the velocities of the electron beams.

## 1. Introduction

Type III bursts have been studied for more than 60 years. Their properties have been analyzed quite well. One of them is the drift of the bursts from high to low frequencies (Suzuki and Dulk, 1985). Within the frame of the plasma mechanism of radio emission, this means that a radio-emission source moves in the solar corona from regions of high to low densities. This mechanism causes radio emissions at the local plasma frequency (fundamental radio emission) or at twice the local frequency (harmonic radio emission) (Ginzburg and Zhelezniakov, 1958). Thus, the radio emission at lower frequency reaches an observer later than the radio emission radiated by the same source at a higher frequency while the radio source propagates outward from the Sun. As a consequence, the frequency-drift rates of type III bursts are negative. Drift rates decrease with decreasing frequency (Suzuki and Dulk, 1985). In the meter range the drift rates are about several hundred MHz s$^{-1}$ (Suzuki and Dulk, 1985), in the decameter range they are equal to some MHz s$^{-1}$ (Abranin, Bazelyan, and Tsybko, 1990; Melnik et al., 2005). This dependence of the drift rate is defined by the radial dependence of the coronal density. It is considered that sources of type III bursts are electron beams moving at a constant velocity (about 0.3c, where c is the speed of light) for large distances from the Sun, up to the Earth orbit and even farther. In the decimeter band, type III bursts are sometimes observed with abnormal positive drift rates (see, for example, Kundu et al., 1961; Elgaroy, 1980; Ma, Xie, and Wang, 2006). These bursts are interpreted as electron beams that propagate in highly inhomogeneous coronal structures, which can be observed near the Sun. It is commonly considered that electron beams moving toward the Sun generate type III bursts with positive drift rates. The bursts with a shape resembling the letter C (Guang-Li et al., 1998; Ma et al., 2008) are interpreted as follows: It is considered in this case that at some heights electrons are accelerated toward and away from the Sun simultaneously. That is, electrons that propagate toward the Sun yield type III bursts with positive drift rates, electrons propagating outward from the Sun generate type III bursts with negative drift rates. There are also so-called inverted U-bursts. Along the tracks of these bursts drift rates also alter their sign. They are interpreted as radio emission from electron beams that propagate along magnetic loops: when electrons move outward from the Sun, they cause radio emission with negative drift rates, when they move toward the Sun, they generate radio emission with positive drift rates (Suzuki and Dulk, 1985). In addition to the idea that a positive drift rate is connected with the motion of the radio source in the direction of increasing density, there is an alternative viewpoint proposed by Chernov (1990). He showed that the sign of the drift rate also depends on the source velocity.

If this velocity is low, the drift rate is negative. At high source velocities this sign is positive. Decameter type III bursts typically have drift rates of 2 – 4 MHz s$^{-1}$ (Abranin, Bazelyan, and Tsybko, 1990; Melnik et al., 2005). But some type III bursts were observed with much higher drift rates of up to 40 MHz s$^{-1}$ (Melnik et al., 2008). Using the equation for the frequency drift rate, $df/dt = (df/dn) \cdot (dn/dr) \cdot v_b$, where n(r) is the plasma density and $v_b$ is the electron beam velocity, one can conclude that these fast type III bursts must be generated by superluminous electrons. To overcome this problem, Melnik *et al.* (2008) took into account the fact that in the vicinity of the radio emission source, the group velocity of the electromagnetic waves can be significantly lower than the speed of light and can be similar to the electron velocity. The comparatively low radio propagation local speed might introduce positive or infinite drift rates. This has been verified by numerical simulations that took into account an actual spatial size of electron beams, their expansion, and direction of propaga- tion (Rutkevych and Melnik, 2012). The drift rates are positive at higher electron velocities when the electron beams move faster than the electromagnetic waves. Type III bursts with positive and negative drift rates were recently observed at frequencies 8 – 41 MHz (Melnik *et al.*, 2013).

In this paper we discuss the properties of these bursts. We show that their parameters, such as drift rates, durations, and radio fluxes, are not different from those of ordinary type III bursts. Therefore it can be expected that type III bursts with alternating drift-rate signs are generated by the same electron beams as normal type III bursts. The only difference is that the change in drift-rate sign seems to be connected with the properties of the plasma through which the electrons propagate. In the regions where the plasma temperature is lower than some given value, the group velocity of the radiated radio emission is lower than the electron velocity, and as a result the electromagnetic waves travel behind the electrons. Consequently, the radio emission at lower frequencies reaches an observer earlier than the radio emission at higher frequencies, and correspondingly, type III bursts have positive drift rates.

## 2. Observations

### 2.1. Instrumentation

Type III bursts with an alternating drift-rate sign were observed by the radio telescopes *Ukrainian T-shaped Radio telescope* (UTR-2, Kharkov, Ukraine: Braude *et al.*, 1978), *Ukrainian Radio telescope of the Academy of Sciences* (URAN-2, Poltava, Ukraine: Megn *et al.*, 2003; Brazhenko *et al.*, 2005), and the *Nançay Decameter Array*, (NDA, Nanç, France: Boischot *et al.*, 1980) during the summer observation campaign in 2012 (Melnik *et al.*, 2013). At UTR-2 only four sections of the radio telescope were used. The two Ukrainian radio telescopes were equipped with a digital spectral polarimeter of z-modification (DSPz) (Ryabov *et al.*, 2010). They enable the recording of radio emission in the continuous frequency band 8 – 32 MHz. The frequency and time resolution of the DSPz used here were 4 kHz and 100 ms. Weber *et al.* (2005) developed the spectrometer that was used for these observations at NDA with a frequency of 12 kHz and time resolution of 37 ms.

### 2.2. Event on June 3, 2012

A group of type III bursts with a changing frequency of their drift rates was recorded on June 3, 2012 during simultaneous observations with the radio telescopes UTR-2, URAN-2, and the NDA (Figure 1a). The most remarkable type III burst was the first burst, which started at 12:04:48 UT in 40 MHz (Figures 1b and c).

This type III burst was observed from 13 MHz to 40 MHz (Figure 1). In the frequency band 28 – 40 MHz the NDA observed a positive drift rate in the forefront of 21.5 MHz s$^{-1}$. According to the UTR-2 and URAN-2 data, the drift rate of the burst forefront is 20 MHz s$^{-1}$ at frequencies 28 – 32 MHz. However, in the frequency range 13 – 28 MHz this drift rate was observed to be $-1.4$ MHz s$^{-1}$. Thus this burst had a positive drift rate at high frequencies and a negative drift rate at lower frequencies. Figures 1b and c show the time evolution of the flux peak of this burst. The burst has a negative drift rate in the frequency bands 32 – 34 MHz, 36 – 38 MHz, and 39 – 40 MHz and a positive drift rate in the frequency bands 29 – 32 MHz, 34 – 36 MHz, and 38 – 39 MHz; that is, the drift rate changes sign three times. The average positive drift rates are 6 MHz s$^{-1}$, 2.5 MHz s$^{-1}$ and 0.7 MHz s$^{-1}$ in the corresponding frequency bands. According to UTR-2 data (Figure 1b), there is also a positive drift rate from 29 MHz to 32 MHz (the drift rate is about 3 MHz s$^{-1}$). At the same time, the drift rate is practically infinite at frequencies 19 – 22 MHz and negative at other frequencies with an average drift rate of about $-1$ MHz s$^{-1}$. This type III burst changes the sign of its drift rate several times.

This type III burst has a duration of about 5 s and a flux of 5 s.f.u. at 41 MHz. At 29 MHz its duration and flux are 9 s and 50 s.f.u. The flux of this burst is practically constant from 29 MHz to 12 MHz, but its duration increases up to 21 s.

### 2.3. Event on August 25, 2012

Another type III burst with alternating drift rate signs was observed on August 25, 2012 (Figure 2a). The drift rate of its forefront was $-3.6$ MHz s$^{-1}$ and 0.95 MHz s$^{-1}$ in the frequency ranges 18 – 32 MHz and 11 – 18 MHz. In the frequency band 21 – 33 MHz the drift rate of its peak changed sign several times with an average negative drift rate of about $-1.7$ MHz s$^{-1}$ (Figure 2b). At lower frequencies the drift rate also changed sign several times, but now the average drift rate is positive, 1.2 MHz s$^{-1}$. This type III burst together with the preceding one resemble an inverted U-burst (Figure 2a). However, we consider this situation to be accidental because all observed decameter U-bursts have a descending part that is essentially fainter than the ascending one (see, for example, Dorovsky *et al.*, 2010; Dorovskyy *et al.*, 2010). Normal U-bursts have descending parts with positive drift rates. This means that the electron beam responsible for the descending part propagates toward the Sun. The discussed burst changes the drift-rate sign. Clearly, this needs to be explained even if this burst is considered as the descending part of a U-burst.

This type III burst has a duration of about 13 s at 32 MHz and 16 s at 13 MHz. Its flux decreases with frequency from $10^3$ s.f.u. at 13 MHz to 10 s.f.u. at 32 MHz.

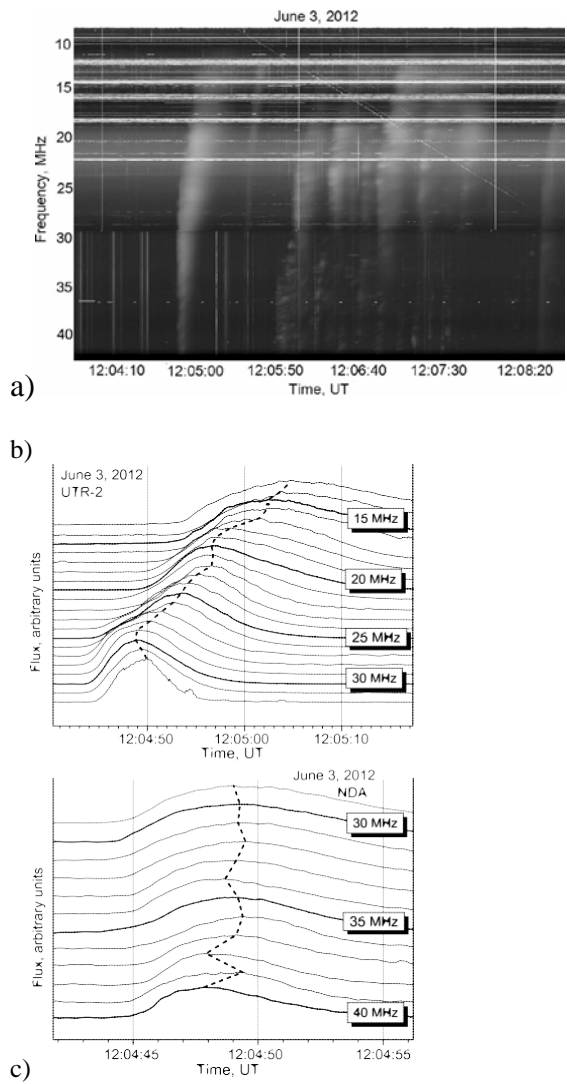

**Figure 1.** (a) The group of type III bursts with changing drift-rate sign recorded with UTR-2 (8 – 29 MHz) and NDA (29 – 41 MHz) on June 3, 2012. (b) and (c): profiles of the flux density at 13 – 32 MHz and 29 – 40 MHz (dashed curves show the peak fluxes at different frequencies).

2.4. Event on August 26, 2012

The next type III burst that exhibited an alternating drift-rate sign was observed on August 26, 2012 (Figure 3a). Its forefront drifts from low to high frequencies with a positive drift rate

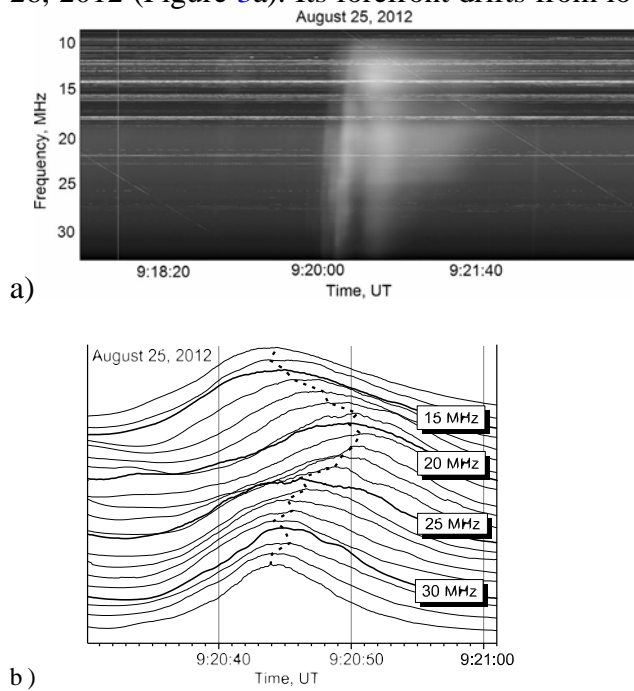

**Figure 2.** (a) Dynamic spectrum of the type III burst with changing drift-rate sign on August 25, 2012 at 09:20:44 UT at 33 MHz (URAN-2 data). (b) The time profiles of this type III burst at different frequencies.

This type III burst was the most intense and longest. Its fluxes varied from 50 s.f.u. to $10^4$ s.f.u. with durations from 25 s to 38 s at 32 MHz and 14 MHz.

The durations of all the analyzed bursts slightly exceed those of normal decameter type of 5.1 MHz s$^{-1}$ . Drifts of the burst peak are observed with a rate of 0.44 MHz s$^{-1}$ from 31 MHz to 27 MHz, with an infinite drift rate from 27 MHz to 22 MHz and a drift rate of 0.9 MHz s$^{-1}$ from 19 MHz to 13 MHz (Figure 3b). The drift rate of this type III burst is negative only in the frequency bands 13 – 14 MHz, 21 – 22 MHz, and 31 – 33 MHz, and equals 0.7 MHz s$^{-1}$ , 1 MHz s$^{-1}$ and 0.7 MHz s$^{-1}$ , respectively. As in the previous case, this type III burst can be considered as the descending part of a U-burst with a flux approximately 10 times higher than that of the ascending part at all frequencies. Therefore, we assume that this burst and the preceding one are two different type III bursts instead of one U-burst. We again note that this burst has a changing drift-rate sign.

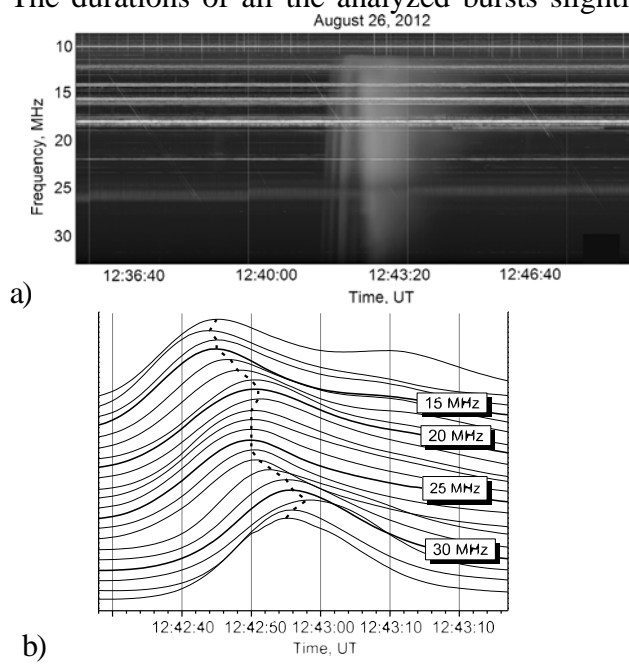

**Figure 3.** (a) The type III burst (12:42:55 UT at 33 MHz) observed with URAN-2 on August 26, 2012 and (b) its time profiles.

III bursts. Their fluxes are similar to the fluxes of standard type III bursts, except for that of the third one, which can be associated with intense type III bursts (Melnik *et al.*, 2011). The frequency drift-rates of these bursts do not differ from those of normal type III bursts except for the alternating signs, therefore we conclude that the change of drift-rate sign of decameter type III bursts cannot be connected with the parameters of type III electrons, but with the properties of the plasma through which these beams propagate, for example, its temperature. We discuss this possibility in Section 3.

### 3. Discussion

We have found that some decameter type III bursts changed the sign of their drift rates while drifting from high to low frequencies in the dynamic spectrum. This type of bursts is frequently observed in the decimeter band (Guang-Li *et al.*, 1998; Ma *et al.*, 2008).

If these decameter bursts are similar in shape to the letter C, in other words, if there are two tracks for the radio emission, one that drifts to high frequencies and another one to low frequencies from the same point, they can be interpreted as being produced by two counter-streaming beams. This interpretation for decameter type III bursts appears to be hardly probable because one needs to assume that the acceleration region must be as high as $1.5 – 2.5$ R$_0$ . In addition, it is not clear how reversed C-bursts could be explained, *i.e.* when the burst has a negative drift rate at high frequencies and then the drift rate sign changes to positive at lower frequencies. It is even more difficult to understand when the burst alters the drift-rate sign more than once.

In our opinion, a possible explanation for a change of drift rate might be the propagation of the radio emission that is generated by fast electrons in a coronal plasma. This interpretation is based on previously stated ideas (Chernov, 1990; Ledenev, 2000; Melnik *et al.*, 2008).

The frequency drift-rate of a type III burst between frequencies $f_1$ and $f_2$ is equal to the ratio of frequency difference, $f_1 - f_2$ , to the difference of radio emission arrival times at Earth, $t_1$ and $t_2$ (Figure 4). As was shown by Melnik *et al.* (2008), this time difference $t_1 - t_2$ is mainly given by the times needed for the electron beam and the electromagnetic wave to travel between the heights $r_1$ and $r_2$, at which the emissions at frequencies $f_1 = f(r_1)$ and $f_2 = f(r_2)$ were generated. Because the group velocity of the fundamental radio emission where it is generated may be much lower than the speed of light, the time difference can be both positive and negative, depending on electron beam velocity and the group velocity, and this can define the drift-rate sign. Following Melnik *et al.* (2008), a source moving with a constant velocity $v_S$ at an angle $\alpha$ to the line of sight (Figure 4) generates radio emission at frequency $f$ with the frequency drift rate

$$\frac{df}{dt} \approx \frac{df}{dn}\frac{dn}{dr}\frac{v_s v_{gr}(r)}{v_{gr}(r) - v_s \cos\alpha} \tag{1}$$

where $v_{gr}(r)$ is the group velocity of the radio emission, and $n(r)$ is the plasma density at the distance $r$ from the Sun. When obtaining Equation (1), we assume that $v_{gr}(r_1) = v_{gr}(r_2) = v_{gr}(r)$ is satisfied if the distance between points $r_1$ and $r_2$ is small compared with the sizes of the inhomogeneities that lead to the observed changes of type III dynamic spectra. We also consider that the difference between travel times from points $r_2$ and $r_1$ to an observer is mainly defined by the distance $r_2 - r_1 \approx \delta r \cdot \cos\alpha$ crossed by the electromagnetic wave (Figure 4). For fundamental radio emission the frequency $f$ equals the local plasma frequency $f = f_{pe} = \frac{\omega_{pe}}{2\pi} = \sqrt{\frac{e^2 n}{\pi m}}$ ($e$ is the electron charge and $m$ its mass), and in this case, $df/dn = f/2n$. The source velocity $v_S$ is assumed to be equal to the velocity of fast electrons $v_0$ (Melnik *et al.*, 2008). This view is valid under the assumption that the Langmuir waves generated by fast electrons do not affect their propagation. Nevertheless, many authors (Ryutov and Sagdeev, 1970; Takakura and Shibahashi, 1976; Magelssen and Smith, 1977; Melrose, 1990) showed that this influence is important. It was shown (Melnik, Lapshin, and Kontar, 1999; Kontar, 2001) analytically and numerically that the interaction between Langmuir waves and fast electrons led to the formation of a so-called beam-plasma structure that moved with the velocity $v_S = v_0/2$ (where $v_0$ was the maximum velocity of fast electrons) when propagating through both homogeneous and inhomogeneous plasmas. This beam-plasma structure can be a source of type III bursts (Melnik and Kontar, 2003). In this article, we also assume that these beam-plasma structures are sources of type III bursts and thus the source velocity $v_S = v_0/2$ in contrast to the condition $v_S = v_0$ used in the article (Melnik *et al.*, 2008). If this is true, the highest spectral energy density of Langmuir waves in the beam-plasma structure will be at the wave number $k_{l,0} \approx \omega_{pe}/v_0$ (Melnik, Lapshin, and Kontar, 1999; Kontar, 2001). In the plasma model of radio emission (Ginzburg and Zheleznyakov, 1958) the fundamental radio emission (the electromagnetic wave $t$) appears because of the processes $l + i = t + i$ because of the scattering of Langmuir waves $l$ by ions $i$. In these processes the electromagnetic wave frequency equals the Langmuir wave frequency $\omega_t = \sqrt{\omega_{pe}^2 + k^2 c^2} = \omega_l = \sqrt{\omega_{pe}^2 + 3k_l^2 v_{Te}^2}$ (here $k_t$ is the wave number of electromagnetic waves, $v_{Te}$ is the thermal velocity of electrons). From this equation, we conclude that electromagnetic waves concentrate mainly at wave numbers $k_{t,o} \approx \sqrt{3}k_{l,o} v_{Te}/c$ and their group velocity

$$v_{gr} = \left.\frac{d\omega_t}{dk_t}\right|_{k_{t,0}} = k_{t,0}\frac{c^2}{\omega_t} \tag{2}$$

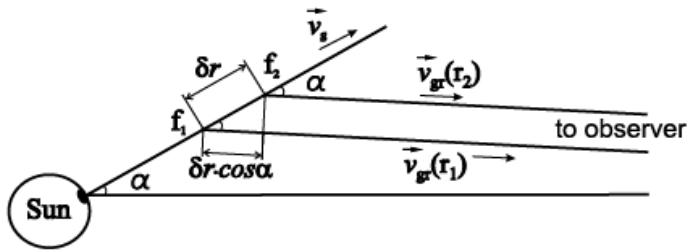

**Figure 4.** Type III exciter propagating at an angle $\alpha$ to the line of sight.

Then the frequency drift rate in Equation (1) is positive for

$$v_{gr} < v_s \cos\alpha \qquad (3)$$

if electrons move outward from the Sun, *i.e.* dn/dr<0 taking into account Equation (2), the condition in Equation (3) can be rewritten as

$$v_0 > v_0^* = \sqrt{\frac{2\sqrt{3}v_{Te}c}{\cos\alpha}} \qquad (4)$$

For electrons moving toward the observer ($\alpha = 0$) we find for the critical velocity $v_0^*$ that

$$v_0^* = \sqrt{2\sqrt{3}v_{Te}c} \qquad (5)$$

which gives $v^* = 7 \cdot 10^9$ cm s$^{-1}$ for plasma temperature $T_e = 2 \cdot 10^6$ K. If we assume that the source velocity remains constant while propagating through the coronal plasma, the observed type III bursts with changing drift-rate signs can be understood in the following way: Fast electrons with velocities lower than $v_0$ generate type III bursts with negative drift rates. When they enter the regions in which the plasma temperature is low enough for $v_0 > v_0^*$, the fast electrons generate type III bursts with a positive drift rate. If later they enter high-temperature regions in which $v_0 < v^*$, the type III bursts will again display negative drift rates. Thus the observation of type III bursts with a change in the drift-rate sign means that the corresponding electron beams pass through regions of different temperatures.

A type III burst was observed on June 3, 2012, changed the sign of its drift rate several times between 28 and 41 MHz and had a negative drift rate at frequencies lower than 28 MHz. To find the correspondence between these frequencies and heliocentric heights we applied the Newkirk model (Newkirk, 1961) for the coronal plasma because it is probable that electron beams responsible for type III bursts propagate through plasma above active regions. Then, we obtained that between heights $r_{41} = 1.6R_\odot$ to $r_{28} = 1.82R_\odot$ (the numerical subscript shows the local plasma frequency in MHz at the given height in the Newkirk model) there were some regions where the critical velocity $v^*$ was either higher or lower than the electron velocity $v_0$. Above $r_{28} = 1.82R_\odot$ the critical velocity remained higher than the electron beam velocity, and the corresponding type III burst had a negative drift rate.

For the burst observed on August 25, 2012, the electrons generated a type III burst with an alternating drift-rate sign from the height $r_{33} = 1.72R_\odot$ to the height $r_{19} = 2.13R_\odot$ (see Figure 2b). This means that there are regions of different temperatures in which the electron velocity $v_0$ is successively higher and lower than the critical velocity $v^*$. This leads to a successive alteration of the drift-rate sign. Starting from the height $r_{19} = 2.13R_\odot$, the plasma temperature decreased and the electron velocity became higher than the critical velocity that gave a positive drift rate.

The situation is different for the type III burst registered on August 26, 2012. The burst had positive drift rates at frequencies 27 – 31 MHz and < 19 MHz (see Figure 3b). This means that at heights from $r_{31} = 1.76R_\odot$ to $r_{27} = 1.85R_\odot$ and starting from $r_{19} = 2.13R_\odot$ and higher there were regions in which the critical velocities were lower than the electron velocity because of a reduced temperature. At heights from $r_{27} = 1.85R_\odot$ to $r_{22} = 2R_\odot$ the plasma temperature was such that $v_0 = v^*$ was fulfilled, and as a consequence, the drift rate of the burst was infinite.

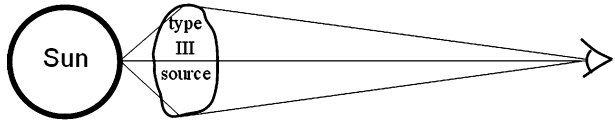

**Figure 5.** Scheme of type III bursts at decameter wavelengths for electron propagation toward the observer.

Our analysis applies to point-like sources of radio emission. In fact, it is known (Abranin *et al.*, 1976) that the region from which decameter type III radio emission comes is larger than the solar radius. This means that the obtained results need to be formulated more exactly. We consider the most favorable configuration for observing the change of the drift rate when electrons move toward an observer, as shown in Figure 5. If the beam size is large enough, the effective critical velocity will be higher than that given by Equation (5) because of a contribution from electrons moving at some angle $\alpha$ to the observer's direction for which the critical velocity is defined by Equation (4). For $\alpha = 45^o$ the critical velocity from Equaion (4) is 1.4 times higher than that for $\alpha = 0$; therefore the effective critical velocity for the type III burst generated by the large beam will be between $\sqrt{2\sqrt{3}v_{Te}c}$ and $1.4\sqrt{2\sqrt{3}v_{Te}c}$.

A detailed numerical study of the propagation of electron beams with different sizes will be developed in the future.

### 4. Conclusion

In addition to normal type III bursts (Abranin, Bazelyan, and Tsybko, 1990; Melnik *et al.*, 2005), fast type III bursts (Melnik *et al.*, 2008) and type III bursts with a changing sign of their drift rates are observed in the decameter range (Melnik *et al.*, 2013). In this wave band such bursts are rare, while in the decimeter range (Guang-Li *et al.*, 1998; Ma *et al.*, 2008) such type III bursts are common. The nature of these phenomena in the decimeter and decameter ranges can be different. Even assuming that decimeter type III bursts with changing drift-rate signs result from the simultaneous propagation of fast electrons toward and away from the Sun, the explanation of decameter type III bursts encounters difficulties. Because both fluxes and drift rates of these unusual decameter type III bursts are similar to those for standard type III bursts, we assumed that the change of the drift-rate sign in these bursts is connected with the properties of the plasma through which the electrons propagate.

If the velocity of the source is lower than the group velocity of the type III radio emission, bursts with negative drift rates are observed. Otherwise, type III bursts have positive drift rates. When these velocities are approximately equal, fast type III bursts are recorded. Thus, from this point of view, all types of decameter type III bursts – standard, fast and with a changing sign of drift rates – can be understood. Moreover, because of the critical velocity for sign change, which is defined by the plasma temperature, we can determine the heights of regions with different temperatures.

**Acknowledgements** The work was partially performed in the frame of the European FP7 project SOLSPANET (FP7-PEOPLE-2010-IRSES-269299).
We thank A.E. Hillaris for useful comments that helped us to improve our article.